\documentstyle[pra,aps,psfig]{revtex}
\begin{document}
\draft
\twocolumn[\hsize\textwidth\columnwidth\hsize\csname @twocolumnfalse\endcsname
\title{Bose condensates in a harmonic trap near the critical temperature}
\author{T. Bergeman,$^1$, D. L. Feder,$^{2,3}$ N. L. Balazs,$^1$ and
B. I. Schneider$^4$}
\address{$^1$Physics Department, SUNY, Stony Brook, NY 11794-3800}
\address{$^2$University of Oxford, Parks Road, Oxford OX1 3PU, U. K.}
\address{$^3$National Institute of Standards and Technology, Gaithersburg, MD
20899-8410}
\address{$^4$National Science Foundation, Arlington, Virginia 22230}
\date{\today}
\maketitle
\begin{abstract}
The mean-field properties of finite-temperature Bose-Einstein gases confined
in spherically symmetric harmonic traps are surveyed numerically. The
solutions of the Gross-Pitaevskii (GP) and Hartree-Fock-Bogoliubov (HFB)
equations for the condensate and low-lying quasiparticle excitations are
calculated self-consistently using the discrete variable representation, while
the most high-lying states are obtained with a local density approximation.
Consistency of the theory for temperatures through the Bose condensation point
$T_c$ requires that the thermodynamic chemical potential differ from the
eigenvalue of the GP equation; the appropriate modifications lead to results
that are continuous as a function of the particle interactions. The HFB
equations are made gapless either by invoking the Popov approximation or by
renormalizing the particle interactions. The latter approach effectively
reduces the strength of the effective scattering length $a_{\rm sc}$,
increases the number of condensate atoms at each temperature and raises the
value of $T_c$ relative to the Popov approximation. The renormalization
effect increases approximately with the log of the atom number, and is
most pronounced at temperatures near $T_c$. Comparisons with the results of 
quantum Monte Carlo calculations and various local density approximations are
presented, and experimental consequences are discussed.
\end{abstract}
\pacs{03.75Fi,05.30.Jp,05.10.-a}]
\narrowtext

\section{Introduction}
\label{sec:introduction}

Since the first observations of Bose-Einstein condensation (BEC) in dilute
alkali metal atom gases~\cite{Anderson,Davis,Bradley}, experimental
developments have posed many new tests for many-body theory, even though
weakly interacting Bose gases have long been used as a textbook
paradigm~\cite{Huang,LandauLif}. Numerous theoretical approaches have been
employed in order to obtain accurate results for both the ground-state and
non-equilibrium properties of the trapped Bose
systems~\cite{StringariRev,WallsRev,HZG}. However, there have been notable
differences between theoretical results and experimental data on the
excitation frequencies near the transition temperature
$T_{c}$~\cite{Jin,Dodd,HDB,Ketterle}. This problem has inspired the introduction
of a renormalized effective atom-atom interaction~\cite{HDB}.
Recently developed theoretical approaches~\cite{ZGN,BijlsmaRes} that 
incorporate the dynamics of the noncondensate density
but without a renormalized interaction have resulted in excitation
frequencies in closer agreement with experiment. Nevertheless, the
unresolved issues for Bose systems near $T_{c}$ has provided a motivation
for us to examine further the theoretical and numerical methods for modeling
confined Bose gases near $T_{c}$. We have numerically
implemented the most plausible and tractable equilibrium mean-field
theories in order to systematically survey various properties of these
systems.

In this work, we follow the standard mean-field theory~\cite{Griffin96}, with
certain modifications described in detail below. The nonlinear
Gross-Pitaevskii (GP) equation, which includes interactions between the
condensate and the thermal atoms, is solved for a static condensate
containing $N_c$ atoms. The eigenvalue of the GP equation, $\tilde{\mu}$,
is usually identified with the thermodynamic chemical potential $\mu$. The
linear response of the system is represented by the Hartree-Fock-Bogoliubov
(HFB) equations, which yield the quasiparticle energies and amplitudes. These
in turn determine the number of noncondensed atoms $N_T$ as well as various
coherence terms (thermodynamic averages over two or more Bose field
operators). The GP and HFB equations are iterated to self-consistency at a
given temperature $T$, subject to a fixed total number of atoms in the
system $N=N_c+N_T$. As emphasized by Griffin~\cite{Griffin96}, the coherence
terms yield an excitation spectrum that is not gapless: the lowest-energy mode
of the HFB equations has finite energy and does not coincide with the solution
of the GP equation. The HFB-Popov (HFBP) approximation, which neglects these
terms, has been quite successful in describing the properties of the
trapped Bose gases, but is not well-grounded theoretically, and fails to yield
accurate predictions for the low-lying excitations at high
temperatures~\cite{Jin,Ketterle}. In this work, we explore a recently proposed
extension of the HFBP theory that incorporates the coherence terms in a gapless
manner~\cite{Proukakis,HDB}; in addition, we modify the commonly used
identification of the chemical potential with the eigenvalue of the GP
equation.

The identification of the chemical potential with the eigenvalue of the GP
equation is incorrect in general. In the grand canonical ensemble, the
chemical potential is defined as $\mu=\partial E/\partial N$, corresponding to
the energy cost $E$ of adding a particle to the entire system, not only to the
condensate. For a dilute, weakly interacting Bose gas at $T=0$, for which the
population of noncondensed states (the depletion) is negligible, the
identification $\tilde{\mu} = \mu$ is justified. At finite temperatures,
however, the assumption yields results that are discontinuous as a function of
the $s$-wave scattering length $a_{\rm sc}$. To a better approximation, we
find that the chemical potential at finite temperatures is given by the
eigenvalue of the GP equation plus a term that varies inversely with the
number of
condensate atoms. The resulting equations provide an improved description of
these finite systems, yielding observables that are both continuous with
$a_{\rm sc}$ and similar to those obtained using path integral Monte Carlo
techniques~\cite{Holzmann}.

It is presently unclear to what extent many-body effects beyond the mean-field
approximation modify the effective interactions among Bose-condensed atoms in
harmonic traps~\cite{HDB,Proukakis,StoofTM,SMorganT}. For the homogeneous Bose
gas, it is now established both from renormalization group~\cite{BStoofRG} and
perturbation~\cite{ShiGriffin} theories that the many-body T-matrix, or
effective s-wave scattering length $a$, goes to zero at $T_{c}$. The
low-energy, long wavelength limit of the many-body T-matrix has been shown to
be closely related to the coherence term $m_T$~\cite{Proukakis,SMorganT}; this
`anomalous average' represents two-particle correlations and is the Bose
analogue of the superconducting order parameter in interacting Fermi systems.
Renormalizing the interaction using the $m_T$ yields a gapless HFB theory
without having to invoke the Popov approximation~\cite{HDB}, but it remains
uncertain whether the prescription is appropriate for nonuniform systems. The
implications of this theory for trapped Bose condensates are explored 
numerically below, and the results are compared to those obtained with
the Popov approximation and path-integral Monte Carlo methods.

In view of these somewhat conflicting results and unresolved issues, there
is strong motivatation for continued development of numerical methods in
order to implement various models and obtain quantitative
predictions for comparison with experiment. Quantum Monte Carlo
methods~\cite{Holzmann,Ceperley,Krauth} are able to provide accurate results
for certain observable quantities. The computational procedure is lengthy,
however, and is not demonstrably able to yield excitation frequencies since it
typically applies only to equilibrium configurations. Local density
approximations (LDA) are much simpler to apply, but the standard forms fail
near $T_{c}$ and are questionable when the density is so small
that the local collision rate is insufficient to establish local thermodynamic
equilibrium. On the other hand, widely used basis set
techniques are generally unable to represent the large numbers of atoms
in excited states at high-temperatures. Recently, Reidl et al.~\cite{Reidl}
have used (for 2,000 Rb atoms at $T = 0.5T_c$) a hybrid method in which a
sum over discrete quasiparticle states at low energies is supplemented by
an integral over an energy-dependent LDA above some cutoff energy. The
interactions of these two subensembles with each other are expressed by
mean-field potentials that represent the effect of background atoms.
In the present work, the low-lying states are obtained
by solving the HFB equations using the discrete variable representation
(DVR)~\cite{Schneider,Colbert,BayeHeenan} and the cutoff energy is
raised until the results converge to within a stated tolerance.
The techniques employed have enabled the investigation of trapped Bose
gases at finite-temperatures containing a larger number of atoms than
in previous calculations that we are aware of. As a result, the approach
of these systems to the local thermodynamic equilibrium and to the
hydrodynamic limit can be explored.

In Section~\ref{sec:mft}, we outline the GP and HFB equations. We  discuss the
chemical potential and gapless theories in Sections~\ref{sec:mu} and
\ref{sec:gapless}, respectively. Section~\ref{sec:lda} reviews LDA methods
both as alternative approaches for comparison purposes, and the complementary
use for the most energetic atoms. In Section~\ref{sec:numerics}, we discuss
our numerical methods and iteration procedures. Section~\ref{sec:results}
presents results for Bose atoms in a spherically symmetric harmonic trap as a
function of the scaled $s$-wave scattering length, total number of atoms, and
temperature.

\section{Theoretical framework}

\subsection{Thermal sums over quasiparticle states}
\label{sec:mft}

The derivation of mean-field equations for a weakly interacting, dilute
Bose gas has been described in detail
elsewhere~\cite{Fetter72,Fetter96,Fetter98,Griffin96}. The question of the
chemical potential for $T>0$ for thermal sums of quasiparticle states
deserves more thorough discussion, however, and we modify the standard
procedure. In addition, following discussions by the Burnett
{\it et al.}~\cite{Proukakis,HDB,SMorganT}, we treat the anomalous (coherence)
terms $m_{T}$ in a manner that produces a `gapless' theory.

Following the standard approach, we decompose the Bose field operator into a
$c$-number for the condensate, plus an operator representing its fluctuations.
The full many-body Hamiltonian is approximated using mean-field theory,
becoming explicitly number-nonconserving. The grand canonical ensemble is used,
and thus the chemical potential, $\mu$, and temperature, $T$, are the sole
fixed quantities. The generalized Gross-Pitaevskii (GP) equation for the
condensate and coupled Bogoliubov equations for the excited quasiparticle
states are then solved. For a finite number atoms in a harmonic potential,
however, the standard approach yields values for the mean condensate number
$N_c$ that are discontinuous as a function of interaction strength
$a_{\rm sc}$. In our approach, the eigenvalue of the GP equation,
$\tilde{\mu}$, is determined by the mean number of atoms in the
condensate $N_c$. In contrast, the chemical potential, $\mu$, is adjusted so
that the mean {\it total} number of atoms is the desired value. A simple
relationship is found connecting $\tilde{\mu}$, $\mu$, and $N_c$, which is
adapted from the ideal Bose gas case.

The Hamiltonian for an interacting Bose gas in a trap in the grand canonical
ensemble is
\begin{eqnarray}
\hat{H}-\mu\hat{N} &= &
\int d{\bf r} \Big\{ \hat{\psi}^{\dagger}
\left[ -\frac{\hbar^2}{2M}\nabla^2 + V_{\text{ext}} -\mu\right]
\hat{\psi}\nonumber \\
& &\qquad \qquad + \frac{g}{2} \hat{\psi}^{\dagger} \hat{\psi}^{\dagger}
\hat{\psi} \hat{\psi} \Big\},  \label{K}
\end{eqnarray}
where the field operator $\hat{\psi}({\bf r})$ satisfies
$[\hat{\psi}({\bf r}_1),\hat{\psi}^{\dagger}({\bf r}_2)]
= \delta({\bf r}_1 - {\bf r}_2)$. The pseudopotential atom-atom
interaction has been chosen to be
$V({\bf r}_1 -{\bf r}_2)=g\delta({\bf r}_1 -{\bf r}_2)$,
where the coupling constant $g = 4 \pi \hbar^2 a_{sc}/M$ is written in terms
of the scattering length $a_{\rm sc}$ and mass $M$. The harmonic potential is
$V_{\text{ext}}={\case 1/2}M\omega_0^2r^2$ with trapping frequency $\omega_0$ is
assumed to be isotropic.

The Hamiltonian may be rewritten as
\begin{eqnarray}
\hat{H}-\mu\hat{N}&=&H-\tilde{\mu}\hat{N}+\left(\tilde{\mu}-\mu\right)\hat{N}
\nonumber \\
&=&\hat{K}+\left(\tilde{\mu}-\mu\right)\hat{N},
\label{Hmod}
\end{eqnarray}
where, as mentioned above, the Lagrange multiplier $\tilde{\mu}$ is
related to the number of atoms in the condensate. In the following, we choose
to diagonalize the operator $\hat{K}=\hat{H}-\tilde{\mu}\hat{N}$ rather than
the original $\hat{H}-\mu\hat{N}$; both choices must lead to the same
excitation spectrum, though with a temperature-dependent shift of the vacuum
for quasiparticle
excitations. In order to make further progress, the Bose field operator
$\hat{\psi} = \Phi + \hat{\phi}$ is now decomposed into $\Phi$, a $c$-number for
the condensate, and $\hat{\phi}({\bf r})$, which annihilates a thermal atom at
${\bf r}$.  The condensate density is defined by $n_{c} = |\Phi|^{2}$, and the
number of condensate atoms is $N_c=\int d{\bf r}|\Phi({\bf r})|^2$. The
noncondensate (thermal) density $n_T$ and anomalous (coherence) terms $m_T$
and $\tilde{m}_T$ are~\cite{Griffin96}
\begin{equation}
n_{T} = \langle \hat{\phi}^{\dagger} \hat{\phi} \rangle;
\qquad m_{T} = \langle \hat{\phi} \hat{\phi} \rangle;
\qquad \tilde{m}_{T} = \langle \hat{\phi}^{\dagger} \hat{\phi}^{\dagger}
\rangle,
\label{nmdefs}
\end{equation}
where the brackets indicate a thermal average in the grand canonical ensemble,
discussed in more detail below. The mean field approximation
is used to reduce the third and fourth order terms to, respectively, first
and second order in $\hat{\phi},\hat{\phi}^{\dagger}$ so that the Hamiltonian
$\hat{K}$ can be diagonalized, following the procedure normally used for
$\hat{H} - \mu \hat{N}$~\cite{Fetter72,Griffin96}.

Excluding the possibility of aggregate motion and vortices~\cite{Fetter96},
$\Phi$ may be taken to be real. The first order terms (plus third order
terms in mean-field approximation) in
$\hat{K}$ vanish if the equation for the condensate is
taken to be the generalized GP equation:
\begin{equation}
\left[ -\frac{\hbar^2}{2M}\nabla^{2} + V_{\text{ext}} + g[n_c + 2 n_{T} +
\tilde{m}_{T}] \right]\Phi = \tilde{\mu}\Phi.
\label{GPE}
\end{equation}
Note that $\tilde{\mu}$ is the eigenvalue of the GP equation. The part of
$\hat{K}$ that is zeroth order in the excited orbitals is a $c$-number
\begin{eqnarray}
K_{0}&=&\int d{\bf r} \Phi({\bf r}) \Big[ -\frac{\hbar^{2}}{2M} \nabla^{2}
+ V_{ext} - \tilde{\mu}\nonumber \\
& &\qquad \qquad + \frac{g}{2} |\Phi({\bf r})|^{2} \Big]\Phi({\bf r}).
\end{eqnarray}
The terms in $\hat{K}$ that are second order in $\hat{\phi}$ are (in the
mean-field approximation) diagonalized by the canonical transformation
\begin{eqnarray}
\hat{\phi}({\bf r}) = \sum_{j} [u_{j}({\bf r}) \hat{\alpha}_{j} +
v_{j}^{*}({\bf r}) \hat{\alpha}_{j}^{\dagger} ]\nonumber  \\
\hat{\phi}^{\dagger}({\bf r}) = \sum_{j} [u_{j}^{*}({\bf r})
\hat{\alpha}_{j}^{\dagger} + v_{j}({\bf r}) \hat{\alpha}_{j} ],
\label{trans}
\end{eqnarray}
such that $[\hat{\alpha}_{i},\hat{\alpha}^{\dagger}_{j}] = \delta_{i,j}$. The
operator $\hat{K}$ is diagonal to second order in $\hat{\phi}$ if the
quasiparticle amplitudes $u_{j}({\bf r})$ and $v_{j}({\bf r})$ are solutions
of the Bogoliubov coupled equations
\begin{eqnarray}
\hat{{\cal L}} u_{j}({\bf r}) + {\cal Q}({\bf r}) v_{j}({\bf
r})&=&\epsilon_{j} u_{j}({\bf r})\nonumber \\
\hat{{\cal L}} v_{j}({\bf r}) + {\cal Q}({\bf r}) u_{j}({\bf r})&=&
-\epsilon_{j} v_{j}({\bf r}),
\label{Bogs}
\end{eqnarray}
where $\hat{\cal L} = K + V_{\text{ext}} - \tilde{\mu} + 2 g n({\bf r})$,
${\cal Q} = g [n_{c}({\bf r}) + m_{T}({\bf r}) ]$, the total density
is $n({\bf r}) = n_{c}({\bf r}) + n_{T}({\bf r})$. For `gapless'
theories, discussed further below, the $j=0$ `Goldstone mode,' has
the property $\epsilon_{0} = 0$, so that $u_{0}({\bf r}) =
-v_{0}({\bf r}) = \Phi({\bf r})$.  Thus on the $\epsilon_{j}$ energy
scale, the condensate has zero energy, and defines the vacuum for
quasiparticle excitations.

After the substitution, $\hat{\psi} = \Phi + \hat{\phi}$, the number
operator $\hat{N} = \int d{\bf r} \hat{\psi}^{\dagger}({\bf r})
\hat{\psi}({\bf r})$ contains terms, such as $\int d{\bf r} \Phi
\hat{\phi}$, that do not conserve particle number. The Bogoliubov
transformation~(\ref{trans}) and coupled equations~(\ref{Bogs}) introduce
a quasiparticle basis such that terms $\hat{\alpha}^{\dagger}_{j}
\hat{\alpha}^{\dagger}_{j}$ and $\hat{\alpha}_{j} \hat{\alpha}_{j}$ are
eliminated, so {\it quasiparticle} number is conserved \cite{Fetter72}. The
diagonalized Hamiltonian explicitly does not conserve particle number, however;
the operator $\hat{K}$ in the quasiparticle basis does not commute with the
excited particle number operator $\hat{\phi}^{\dag}\hat{\phi}$, which has
contributions from $\hat{\alpha}^{\dag}\hat{\alpha}^{\dag}$ and
$\hat{\alpha}\hat{\alpha}$ terms. In the grand canonical ensemble, only $T$
and $\mu$ are precisely defined, and all observables must be defined in terms
of thermal averages. Each occupation number, including the condensate number,
fluctuates about its mean value
\begin{equation}
\langle\hat{N}_j\rangle\equiv\langle\hat{\alpha}^{\dag}\hat{\alpha}\rangle,
\quad j=0,1,\ldots,
\label{Nj}
\end{equation}
where the explicit definition of the average $\langle\hat{O}\rangle$ is yet
undefined. Similarly, both the eigenvalue $\tilde{\mu}$ of the GP
equation~(\ref{GPE}) and the total energy $\langle E \rangle$ fluctuate about
their mean values.

Inserting the transformation (\ref{trans}) into Eqs.~(\ref{nmdefs}) and
introducing the identification given by Eq.~(\ref{Nj}), the normal and
anomalous densities become
\begin{equation}
n_T({\bf r}) = \sum_{j=1} \left\{ \langle \hat{N}_{j} \rangle
[|u_{j}({\bf r})|^{2} + |v_{j}({\bf r})|^{2}] + |v_{j}({\bf r})|^{2}\right\};
\label{Nex}
\end{equation}
\begin{equation}
m_{T}({\bf r}) = \sum_{j=1} u_{j}({\bf r}) v_{j}^{*}({\bf r})[2 \langle
\hat{N}_{j} \rangle + 1].
\label{anomaly}
\end{equation}
The standard normalization $\int d{\bf r} [|u_{j}({\bf r})|^{2} -
|v_{j}({\bf r})|^{2}] = 1$, yields
\begin{equation}
\int d{\bf r} [|u_{j}({\bf r})|^{2} + |v_{j}({\bf r})|^{2}]\equiv 1 + 2 V_{j},
\end{equation}
where $V_{j}=\int d{\bf r}|v_{j}({\bf r})|^{2}$. The quantities
$V_{j}$ are related to the $T=0$ depletion, which is $\sum_{j=1} V_{j}$.
The relation between the total atom number and the quasiparticle
occupation numbers is therefore
\begin{eqnarray}
\langle \hat{N}\rangle & \equiv & N_c + N_T =
\langle N_0 \rangle + \int d{\bf r}\, n_{T}({\bf r}) \nonumber \\
& = & N_c + \sum_{j=1} [\langle \hat{N}_{j} \rangle
(1 + 2 V_{j}) + V_{j}],
\label{NNCNT}
\end{eqnarray}
where the average number of atoms is written in terms of a contribution from
the condensate and noncondensate (excitations).

The thermal average of the diagonalized Hamiltonian then becomes
\begin{eqnarray}
\langle \hat{H}&-&\mu \hat{N} \rangle = K_{0} + \sum_{j=1}
\epsilon_{j} (\langle\hat{N}_{j} \rangle - V_{j})+
\left(\tilde{\mu}-\mu\right)\langle\hat{N}\rangle \nonumber \\
&=& E_{c} - \mu N_{c}
+ \sum_{j=1} \big\{ \langle \hat{N}_{j} \rangle [\epsilon_{j}
+(\tilde{\mu}-\mu)(1 + 2 V_{j})] \nonumber \\
&&\qquad\qquad\qquad + V_{j}(\tilde{\mu} - \mu - \epsilon_{j}) \big\},
\label{HH}
\end{eqnarray}
where $E_{c} = K_{0}+\tilde{\mu}N_c$ is the total ground state, or
condensate, energy. 

\subsection{Occupation factors and the chemical potential}
\label{sec:mu}

In the Bogoliubov approach, the ensemble is considered to be the sum of a
condensate plus non-interacting quasiparticles. The mean occupation numbers
of the quasiparticle states are to be determined from the grand partition
function,
\begin{equation}
\Omega={\rm Tr}\{\exp{[-\beta(\hat{H}-\mu\hat{N})]}\},
\label{part}
\end{equation}
through the standard identities~\cite{Huang,LandauLif}
\begin{equation}
\langle N \rangle={1\over\beta}
\left({\partial\ln\Omega\over\partial\mu}\right)_{T}; \ \
\langle E \rangle=-\left({\partial\ln\Omega\over\partial\beta}\right)_{\mu,T}.
\label{bardefs}
\end{equation}
Unfortunately, while the diagonalized Hamiltonian is written in terms of 
non-interacting single-quasiparticle energies, the expressions~(\ref{NNCNT})
and (\ref{HH}) involve the thermal averages of particle occupation that we are
now seeking to determine. Furthermore, the factorization that one makes for an
ideal Bose gas is invalid for a gas of interacting Bose atoms because the
quasiparticle energies depend on the occupation numbers, as well as the
reverse. Thus, rigorously, these occupation factors should be calculated
self-consistently, along with Eqs.~(\ref{GPE}) and (\ref{Bogs}), since they
depend on as well as {\it determine} the quasiparticle
eigenvalues~\cite{Gajda}. To do so analytically would be a truly daunting task.
We make several simplying assumptions in order to obtain results, but we
emphasize that these questions merit further study.

In reality, the probabilities $\langle\hat{N}\rangle$ will be peaked at the
most probable values, as discussed below for the condensate. Therefore, when
evaluating the sum over $N_{j}$ in Eq.~(\ref{part}), deviations of $N_{j'}$
from $\langle\hat{N}_{j'}\rangle$ for $j'\neq j$ will not greatly modify the
spectrum of the quasiparticle states. If this is so, a reasonable approximation
is to replace $\langle\hat{N}_{j}\rangle$ by $N_{j}$ when estimating the mean
occupation numbers from the grand partition function. If the dependence of
$\epsilon_{j}$ and $N_{j}$ on $N_{j'} (j \neq j')$ is also neglected, then
$\Omega$ can be factored, and we obtain
\begin{eqnarray}
\langle N_j\rangle&=&{\sum_{N_j}N_j\exp\left\{-\beta\left[\epsilon_j+
(\tilde{\mu}-\mu)(1 + 2 V_{j}) \right]N_j\right\}\over
\sum_{N_j}\exp\left\{-\beta\left[\epsilon_j+(\tilde{\mu}-\mu)(1 + 2 V_{j})
\right]N_j\right\}}\nonumber \\
&\approx&{1\over\exp\left\{\beta\left[\epsilon_j
+(\tilde{\mu}-\mu)(1+ 2V_{j})\right]\right\}-1},\quad\forall j.
\label{occ}
\end{eqnarray}
In order to obtain the result on the second line of Eq.~(\ref{occ}), the
population-dependences of the GP eigenvalue $\tilde{\mu}$ and the quasiparticle
energies $\epsilon_j$ are ignored. At sufficiently low temperatures, the
$\epsilon_j$ for trapped Bose-condensates are relatively insensitive to the
value of $N_c$ and the temperature; indeed, in the Thomas-Fermi (TF) limit,
valid for large condensates, the excitation frequencies at zero temperature
are independent of $N_{c}$.

Neglecting the factors $V_j$, and shifting the energy scale so that
$E_j\equiv\epsilon_j+\tilde{\mu}$, one recovers the more conventional
expression
\begin{equation}
\langle N_j\rangle = {1\over\exp[\beta(E_j-\mu)]-1}.
\label{Njj}
\end{equation}
From this expression (\ref{Njj}) for $j=0$, with $E_0=\tilde{\mu}$, one finds
that the chemical potential $\mu$ and the eigenvalue of the GP equation
$\tilde{\mu}$ are related by the expression
\begin{equation}
\mu=\tilde{\mu}-{1\over\beta}\ln\left(1+\frac{1}{N_c}\right),
\quad N_c>0.
\label{muT}
\end{equation}
For $T=0$ this gives the usual definition $\tilde{\mu} = \mu$, but for $T>0$
there is a correction to $\mu$ that increases as $N_c$ decreases. While this
additional term will not be correct at high temperatures where the condensate
is strongly depleted, it will be shown below that results obtained with this
procedure are continuous functions of $a_{\rm sc}$ at all temperatures, while
with $\mu=\tilde{\mu}$ they are not.

It is difficult to go beyond the above approximations, but we will suggest
possible avenues to proceed in future work.  The major effect omitted is
the dependence of the quasiparticle energies, $E_{j}$ (including
$E_0=\tilde{\mu}$) on $N_c$. One can first consider the condensate term itself.
We assume, for the moment, that factorization of $\Omega$~(\ref{part}) is
valid, and write
\begin{equation}
\Omega = \Omega_{c} \Omega_{T}; \ \ \Omega_{c} =
\sum_{N_c} e^{\beta(\mu N_{c} - E_{c})} \label{Omegac}
\end{equation}
In the Thomas-Fermi approximation (kinetic energy in the GP equation
neglected), one obtains for a spherical condensate~\cite{StringariRev}:
\begin{equation}
\tilde{\mu}_{\rm TF}=\frac{1}{2}\left(\frac{15N_{c}a_{sc}}{a_{0}}\right)^{2/5}
\equiv \gamma N_{c}^{2/5},
\label{TF}
\end{equation}
where the harmonic oscillator length is $a_0=\sqrt{\hbar/M\omega}$. The
following relations follow in the same approximation:
\begin{equation}
E_{c} = \frac{5}{7} \gamma N_{c}^{7/5} = \frac{5}{7} \tilde{\mu}_{\rm TF}N_{c};
\ \ \frac{\partial E_{c}}{\partial N_{c}} = \tilde{\mu}.
\end{equation}
Then from Eq.~(\ref{Omegac}), neglecting $\Omega_{T}$, one can obtain
the mean value of the condensate occupation from
$\langle N_c\rangle=\sum_{N_{c}}N_{c}P(N_{c})/\sum_{N_{c}}P(N_{c})$, where
$P(N_{c})=\exp[\beta(\mu N_c-{\case 5/7}\gamma N_c^{7/5})]$. We have verified
numerically for typical values of $\beta$ and $\mu$ that the mean value
is extremely close to the most probable value $\bar{N}_c$ for which
$P(N_{c})$ is maximum.
Furthermore, an expansion of the exponent in the above expression for
$P(N_{c})$ yields a value for the variance of $N_{c}$, interpreted as the
value of $\sigma$ in such that $P(\bar{N}_{c}\pm\sigma)=(1/e)P(\bar{N}_{c})$.
In the grand canonical ensemble at zero temperature, therefore, one obtains
$\langle\delta N_{c}\rangle=\sqrt{(5/\beta\gamma)}\bar{N}_{c}^{3/10}$, so that
the fractional width of the occupation number distribution decreases as
$\bar{N}_{c}^{-7/10}$. This may be compared with the result of Giorgini et
al., derived from excited state occupation numbers for the canonical ensemble,
$\langle\delta N_c\rangle\sim(T/T_c)N_c^{2/3}$~\cite{GPS}. Either result
confirms that the fluctuations in $N_{c}$ are relatively small for large
$N_{c}$.  One should next consider how the dependence of $\Omega_{T}$
would effect $\langle N_{c} \rangle$ and $\langle \delta N_{c} \rangle$.
This is left for future work.
 
The dependence of the quasiparticle states on the occupation factors reflects
the extensive nature of this finite interacting system; that is, adding a
particle to the many-body system alters both the number and character of the
accessible states. This behavior is similar~\cite{Wu,Bhaduri} to that of a
finite gas of non-interacting particles obeying {\it fractional exclusion
statistics}~\cite{Haldane}, which obey a statistics intermediate between that
of bosons and fermions. The parameter representing the statistics has been
identified with the strength of the delta-function potential for an
interacting trapped Bose gas in two dimensions~\cite{Bhaduri}. Indeed, our
expression~(\ref{muT}) for the thermodynamic chemical potential is similar to
that found for a non-interacting fractional-statistics gas at finite
temperature~\cite{Wu,Bhaduri}. We hope to pursue these issues more fully in
future work.

\subsection{Gapless approximations}
\label{sec:gapless}

We return to the conditions for `gaplessness.' The GP (\ref{GPE}) and
Bogoliubov (\ref{Bogs}) equations together comprise the
`Hartree-Fock-Bogoliubov' (HFB) approximation for a dilute interacting Bose
gas. In this case, one does not obtain $\epsilon_{0}=0$, and the theory is
said to be not gapless (the term has been taken from the homogeneous
situation). In the Popov approximation, gaplessness is ensured by neglecting
the coherence terms $m_{T}$ and $\tilde{m}_{T}$, but the
justification for such an approximation is questionable~\cite{Griffin96}.

In order to convert HFB into a gapless theory and still retain the anomalous
averages, Burnett {\it et al.}~\cite{Proukakis,HDB} have recently proposed
an alternative treatment in which the coupling functions for the condensate
$g_c({\bf r})$ and excited states $g_e({\bf r})$ absorb the pairing
correlations, and thereby take on a spatial dependence. Eq.~(\ref{GPE})
becomes
\begin{equation}
\{ K + V_{\text{ext}} + g_{c} n_c + 2 g_{e} n_{T}  \} \Phi = \tilde{\mu}\Phi,
\label{GPEM}
\end{equation}
and similarly, $\hat{\cal L}$ and ${\cal Q}$ appearing in Eqs.~(\ref{Bogs})
become
\begin{equation}
\hat{\cal L}=K+V_{\rm ext}-\tilde{\mu}+2g_cn_c+2g_en_T;\quad{\cal Q}=g_cn_c.
\end{equation}
In the proposed gapless theories, labeled G1, and G2, the coupling constants
are chosen to be:
\begin{equation}
\left\{g_c; g_e\right\}=\cases{\left\{g_1; g\right\}, & G1 \cr
\left\{g_1; g_1\right\}, & G2},
\label{G123}
\end{equation}
where
\begin{equation}
g_{1}({\bf r}) = g \left[1 + {m_{T}({\bf r})\over n_{c}({\bf r})}\right].
\label{gr}
\end{equation}
The renormalized coupling $g_1$ replaces the two-body T-matrix associated with
binary atomic collisions, which is the scattering length $a_{\rm sc}$ in 
vacuo, by the zero momentum and energy limit of the homogeneous many-body
T-matrix~\cite{Proukakis,HDB,SMorganT}. In the G1 approximation, only the
condensate-condensate and condensate-excited are dressed, while G2 is
motivated by the expectation that all particle interactions should be similar.
Renormalization of the coupling has the additional advantage of removing the
ultraviolet divergence in $m_T$ resulting from high-energy quasiparticle
contributions of Eq.~(\ref{anomaly}) in the T-matrix approximation. In
nonuniform systems such as the trapped Bose gases, however, the value of
$g_{1}({\bf r})$ can diverge in regions near the condensate surface where the
condensate density vanishes more rapidly than the anomalous average. In
practice, this divergence may be eliminated by setting
$g_{1}({\bf r})=g\left[1+m_{T}({\bf r})/\left(n_{c}({\bf r})
+\delta\right)\right]$, where $\delta\approx 10^{-2}$. While the results, 
described in detail below, are found not to depend strongly on the choice of
$\delta$, its existence underlines a deficiency in the theory in its present
form. The consequences of the G1 approximation are not explored in this work.
In the following, the notation $g({\bf r})$ will be used in place of
$g_{1}({\bf r})$ and in distinction to $g$, which is unrenormalized.

\subsection{Local Density Approximation}
\label{sec:lda}

In local density approximation (LDA) schemes, the condensate density
is assumed to be varying sufficiently slowly that the population of
excited states is determined entirely by the local potential and temperature.
The thermal density may then be treated locally as if the interacting
Bose gas were homogeneous. We will discuss three basic LDA schemes,
and several variants.

In the semiclassical approximation to the GP and HFB
equations~\cite{Reidl,GiorginiLTP}, the thermal atom quasiparticle amplitudes
in the Bogoliubov equations~(\ref{Bogs}) become local functions
$u({\bf p},{\bf r})$ and $v({\bf p},{\bf r})$. With the Popov approximation,
one obtains the coupled algebraic equations
\begin{equation}
\pmatrix{{\cal L}({\bf p},{\bf r}) & g n_{c}({\bf r}) \cr -g n_{c}({\bf r})& -{\cal L}({\bf p},{\bf r})}
\pmatrix{ u({\bf p},{\bf r}) \cr v({\bf p},{\bf r})}
= \epsilon({\bf p},{\bf r})
\pmatrix{ u({\bf p},{\bf r}) \cr v({\bf p},{\bf r})},
\end{equation}
where ${\cal L}({\bf p},{\bf r})=p^{2}/2m+V_{\rm ext}({\bf r})-\tilde{\mu}
+2gn({\bf r})$. With the condition
$u({\bf p},{\bf r})^{2} -v({\bf p},{\bf r})^{2} = 1$, the local excitation
energies may be immediately obtained, $\epsilon({\bf p},{\bf r})
= ({\cal L}({\bf p},{\bf r})^{2} - g^{2} n_{c}^{2}({\bf r})^{2})^{1/2}$, and
have the well known linear dispersion. The noncondensate density from
Eq.~(\ref{Nex}) may then be easily found~\cite{Reidl}:
\begin{eqnarray}
n_{T}({\bf r})&=&\int \frac{d^{3}{\bf p}}{(2\pi)^{3}} \left[ \frac{{\cal
L}({\bf p},{\bf r})}{\epsilon({\bf p},{\bf r})}
\left(\langle n({\bf p},{\bf r})\rangle+{1\over 2}\right)-{1\over 2}\right]
\label{LDAnt}
\nonumber \\
& &\quad\Theta\left({\cal L}({\bf p},{\bf r})^{2} - g^{2}n_{c}^{2}({\bf r})
\right),
\end{eqnarray}
where
\begin{equation}
\langle n({\bf p},{\bf r})\rangle={1\over\exp[\beta(\epsilon({\bf p},{\bf r})
+\tilde{\mu}-\mu)]-1},
\label{Occ}
\end{equation}
such that the theta function is unity when the argument is positive, and zero
otherwise. These equations define the Hartree-Fock Bogoliubov Popov LDA, which
we will refer to as the ``BPLDA.''  For G2 calculations, one obtains the
``BGLDA'' by the substituion $g \rightarrow g({\bf r})$ everywhere. Then one
needs
\begin{eqnarray}
m_{T}({\bf r})&=&\int \frac{d^{2}{\bf p}}{(2\pi)^{3}} u({\bf p},{\bf r})
v({\bf p},{\bf r}) [2 \langle n({\bf p},{\bf r}) \rangle +1]\nonumber \\
&=& -g({\bf r})n_{c}({\bf r})\int \frac{d^{3}{\bf p}}{(2\pi)^{3}}
\frac{1}{2\epsilon} \left[2\langle n({\bf p},{\bf r})\rangle +1 \right]
\nonumber \\
& &\qquad\Theta\left({\cal L}({\bf p},{\bf r})^{2}-g^{2}n_{c}^{2}({\bf r})
\right).
\label{LDAmt}
\end{eqnarray}
The integral is not formally convergent, however. Since the anomalous averages
appear only in the context of the G1 and G2 approximations, where the formal
ultraviolet divergence is eliminated, we may safely neglect the $+1$ term
following the $2\langle n({\bf p},{\bf r})\rangle$.

The semiclassical HFBP approximation exhibits a gapless excitation spectrum
only if the condensate is also treated within the LDA, which implies the
TF density:
\begin{eqnarray}
n_{c}({\bf r})&=&\frac{\tilde{\mu}-V_{\rm ext}({\bf r})-2gn_{T}({\bf r})}{g}
\nonumber \\
& &\quad\Theta\left[\tilde{\mu}-V_{\rm ext}({\bf r})-2gn_{T}({\bf r})\right],
\label{TFA}
\end{eqnarray}
The TF approximation is valid in the limit of large $N_c$, where
the energy contribution from the mean-field (Hartree) potential exceeds
that of the kinetic energy. For this reason, Eq.~(\ref{TFA}) is not expected
to be a good representation of the condensate density close to the transition
temperature.

In the regime of small condensate numbers, therefore, it becomes more important
to solve the equations for the condensate and excitations exactly in order to
obtain the low-lying discrete states, as described in the previous section. In
this work, we use the exact GP and HFB equations, but the sum over discrete
states is combined with an energy integral over high-lying states using LDA
functions in the manner described by Reidl {\it et al.}~\cite{Reidl}:
\begin{equation}
n_{T}({\bf r}) = \sum_{j} n_{j}({\bf r}) \Theta(\epsilon_{c} - \epsilon_{j}) +
\int_{\epsilon_c}^{\infty} d\epsilon n_{T}(\epsilon,{\bf r}), \label{sumint}
\end{equation}
where $n_{j}({\bf r})$ is the $j$th term of Eq.~(\ref{Nex}), $\epsilon_c$ is a
low-energy cutoff, and, in the above notation, $n_T(\epsilon,{\bf r})$ has
the form
\begin{eqnarray}
n_{T}(\epsilon,{\bf r})& =& \frac{m^{3/2}}{2^{3/2}\pi}
\left[ 2 \langle n({\bf p,r}) \rangle + 1 - \frac{\epsilon}{{\cal L}}\right] 
\nonumber \\ 
& \times &\left[ {\cal L} - V_{\rm ext} + \tilde{\mu} - 2 g n\right]^{1/2}.
\end{eqnarray}
A similar equation applies to the anomalous average $m_T$. This latter
hybrid procedure is referred to below as the Discrete Quasiparticle Sum
(DQS) approximation, an abbreviation for `discrete
Hartree-Fock-Bogoliubov quasiparticle sum.' Either a Popov or G2 approximation
may be made within the DQS, and these are referred to below as DQSP and DQSG,
respectively.

A simpler LDA may be formulated by treating the local excitations
within the Hartree-Fock approximation, which ignores the linear dispersion at
low energies. The condensate density may again be obtained within the TF
approximation using Eq.~(\ref{TFA}). The thermal density is given by
$n_T({\bf r})=\int{d^3p\over(2\pi)^3}\langle n({\bf p},{\bf r)}\rangle$, where
$\langle n({\bf p},{\bf r})\rangle$ is defined in Eq.~(\ref{Occ}) but with
$\epsilon({\bf p},{\bf r})={\cal L}({\bf p},{\bf r})$. Integration over the
momenta readily yields
\begin{equation}
n_{T}({\bf r})=\frac{1}{\lambda_{T}^{3}} g_{3/2}\left(e^{-\beta(
V_{\text{ext}}({\bf r}) + 2 g n({\bf r})-\mu)} \right),
\label{HFLDA}
\end{equation}
where the thermal de~Broglie wavelength is
$\lambda_{T}=(2\pi\hbar^{2}/mkT)^{1/2}$ and
$g_{\alpha}(z) = \sum_{j=1}^{\infty}z^{j}/j^{\alpha}$. As usual, the chemical
potential $\mu$ is determined by the condition that the total atomic number,
$N = \int d{\bf r}[n_c({\bf r}) + n_{T}({\bf r})]$.  With the TF
expression~(\ref{TFA}) for the condensate, the argument of the $g_{3/2}$
function in Eq.~(\ref{HFLDA}) is always less than unity. If an `exact'
solution for the condensate is used (i.e.\ obtained by solving the GP
equation), the results are generally improved, but as noted below and in
Ref.~\onlinecite{Holzmann}, there is then a range of temperatures 
$T\lesssim T_{c}$ for which the $g_{3/2}$ function given in Eq.~(\ref{HFLDA}) 
diverges, since its argument can become greater than unity.

An even simpler form of the LDA has been formulated~\cite{GiorginiLTP,NSK} in
which the effect of interactions on the excited states is completely ignored.
Assuming a TF form for the ground state, this LDA consists of the
parametrically coupled equations (in view of the other approximations here, in
these equations we ignore the distinction between $\tilde{\mu}$ and $\mu$):
\begin{eqnarray}
n_c({\bf r}) = \frac{\tilde{\mu}-V_{\rm ext}({\bf r})}{g}\Theta[\tilde{\mu}
-V_{\text{ext}}({\bf r}) ] \\
n_{T}({\bf r}) =\frac{1}{\lambda_{T}^{3}} g_{3/2}\left(e^{-\beta\left|
V_{\text{ext}}({\bf r}) - \tilde{\mu}\right|} \right).
\label{SILDA}
\end{eqnarray}
In this approximation, the interaction enters only via the
chemical potential in the TF equation, which is a function of $a_{\rm sc}$ and
condensate number. For a spherical condensate,
$\tilde{\mu}_{\rm TF}={\case 1/2}(15 N_c a_{\rm sc}/a_{0})^{2/5}\hbar\omega_0$,
where $a_0=\sqrt{\hbar/M\omega_0}$ is the bare oscillator length. 

It is shown in Ref.~\onlinecite{NSK} that a low order expansion of
Eq.~(\ref{SILDA}) yields the following expression for $N_{0}/N$:
\begin{equation}
\frac{N_{c}}{N} = 1 - \left( \frac{T}{T_{c}^{0}} \right)^{3} - \eta
\frac{\zeta(2)}{\zeta(3)} \left( \frac{T}{T_{c}^{0}} \right)^{2}
\left( \frac{N_{c}}{N} \right)^{2/5},
\label{SILDAN}
\end{equation}
where $\zeta(n)$ is the Riemann zeta function,
$\eta = \tilde{\mu}_{\rm TF}/k_BT_{c}^{0}\approx{\case1/2}\zeta(3)^{1/3}
(15N^{1/6}a_{\rm sc}/a_0)^{2/5}$, and the critical temperature for $N$ ideal
Bose atoms in a harmonic trap is given by \cite{Grossmann,Balazs}
\begin{equation}
k_{B}T_{c}^0/\hbar \omega_0 = 0.9405 N^{1/3} - 0.6842 + 0.50 N^{-1/3}
\label{Tc}
\end{equation}
Equation (\ref{SILDAN}) is solved iteratively for $N_{c}/N$.

\subsection{Ideal Bose gas}

Some of the plots given below contain results for ideal non-interacting
Bose atoms ($a_{\rm sc}=0$) in a harmonic trap. The results given for $N_{c}$
were obtained from sums over the occupation numbers as given in
Eq.~(\ref{Nj}), with $d_{j} = 2 \ell_{j}+1$, $E_{j} = \hbar \omega(\ell +
2n_{j}+3/2)$. The chemical potential $\mu$ was adjusted to satisfy the
condition $N = \sum_{j=0} \langle N_{j} \rangle$. An alternative expression
can be obtained from the density distribution given by
Chou {\it et al}.~\cite{Chou}:
\begin{eqnarray}
N&=&\frac{z_{1}}{1 - z_{1}}\nonumber \\
&+&\sum_{l=1}^{\infty} z_{1}^{l}
\left\{[(1 - e^{2l\beta})(\tanh(\beta l/2)]^{-3/2} - 1 \right\},
\label{Nideal}
\end{eqnarray}
where $z_{1} = e^{\beta (\mu - 3/2)}$.  This expression requires fewer terms
than the aforementioned procedure, and gives identical results for
temperatures up to about $0.9T_c$.

\section{Computational techniques}
\label{sec:numerics}

With a spherically symmetric trapping potential, all observables may be
decomposed into functions of radius $r$ and spherical harmonics
${\cal Y}_l^m(\theta,\phi)$. The GP and Bogoliubov equations then become
one-dimensional in $r$; the ground state is assumed to have
$(\ell,m)=(0,0)$, while the excitations obtained using the Bogoliubov equations
are $2 \ell +1$ degenerate. Both equations are solved using the discrete
variable representation (DVR), a computationally efficient approach for the
trapped interacting Bose gases that has been recently described in
detail~\cite{Schneider}.

We have used two variants of the DVR approach: an equidistant mesh array
derived from sine functions as discussed by Colbert and Miller~\cite{Colbert},
and a mesh based on Gaussian quadrature, using the zeros of associated
Laguerre polynomials $L_{N_L}^{\alpha}(r)$, where $N_L$ is the order of the
the quadrature and $\alpha=2$ for a spherical condensate~\cite{BayeHeenan}.
The latter DVR has the advantage of having a fine mesh for small $r$ where
the condensate density is non-zero, and a more coarse mesh at larger distances
where the thermal distribution varies smoothly. Though the condensate and
excited orbitals are computed on the physical grid, the matrix elements of the
operators are represented by Laguerre polynomials up to the order defining the
Gauss quadrature $N_L$, which in the present calculations range from 1,000 to
2,800; matrix elements of the kinetic energy are computed from expressions
given in Ref.~\onlinecite{BayeHeenan}. Increasing the value of $N_L$ increases
the accuracy of high-lying states, allowing for a larger cutoff energy
$\epsilon_c$ at which the discrete sums are terminated, and a smaller number of
atoms in the LDA integrals. Since high-order polynomials extend far beyond
values of $R_{\rm max}\lesssim 50a_0$ relevant to trapped condensates,
the number of spatial grid points required can be limited to just
$N_g\sim 200$ for all values of $N_L$.

Implementation of the above mean-field theory requires a stable and efficient
iteration procedure to solve the GP and Bogoliubov equations for a given total
number of atoms $N$ and temperature $T$. In our approach, the functions
$n_c(r)=\Phi^2(r), n_{T}(r)$, and $m_{T}(r)$ are
calculated self-consistently using Eqs.~(\ref{GPE}) and
(\ref{Bogs})-(\ref{anomaly}), supplemented by the LDA expressions for states
above the cutoff $\epsilon_c$, for fixed $N_c$ and $T$; the chemical potential
$\mu$ is determined by Eq.~(\ref{muT}). Because this iteration procedure is
especially delicate near $T_{c}$, yet is crucial for the results presented,
we give a few more details.

We emphasize that the convergence criterion must consider the spatial
distribution functions $n_{c}({\bf r})$ and $n_{T}({\bf r})$ rather than
simply the aggregate values, $N_{c}$ and $N_{T}$. The iterative procedure can
be decomposed into three separate levels of self-consistency, subject to the
minimization of the `Error':
\begin{equation}
{\rm Error} = \int d{\bf r}\big[|n_{c}^{\rm out}({\bf r})
- n_{c}^{\rm in}({\bf r})|
+ |n_{T}^{\rm out}({\bf r}) - n_{T}^{\rm in}({\bf r})|\big].
\label{Error}
\end{equation}
The `in' and `out' functions are the input and output of the combined GP and
HFB equations plus the high energy LDA integral. Normally, the Error
diminishes (though not necessarily monotonically) through level 1 iterations,
in which the output functions are fed back into the GP, HFB and high energy
LDA equations. In this level, the condensate number $N_c$ is held constant
while the condensate density (normalized to unity) is allowed to vary. When
the Error reaches some predetermined tolerance, level 2 iterations begin and
$N_{c}$ is adjusted to approach the condition that $N_{c} + N_{T} = N$. The
first level 2 adjustment from the converged level one iterations is based on
a simple proportionality between $N$ and $N_{c}$.  Subsequent level 2
adjustments are based on a linear relation between $N_{c}$ and $N$, where the
parameters are obtained from the last two level 2 iterations. After $N_c+N_T$
has converged to $N$ to the desired tolerance, level 3 iterations proceed, in
which iteration levels 1 and 2 are repeated with successively larger number of
Laguerre functions $N_L$ and mesh points $N_g$. These three levels of
iteration typically achieve accuracies for the condensate number $N_c$ of a
few atoms. While this accuracy is beyond what is accessible to current
experiments, it permits the comparison of different theoretical models.

\begin{figure}
\psfig{figure=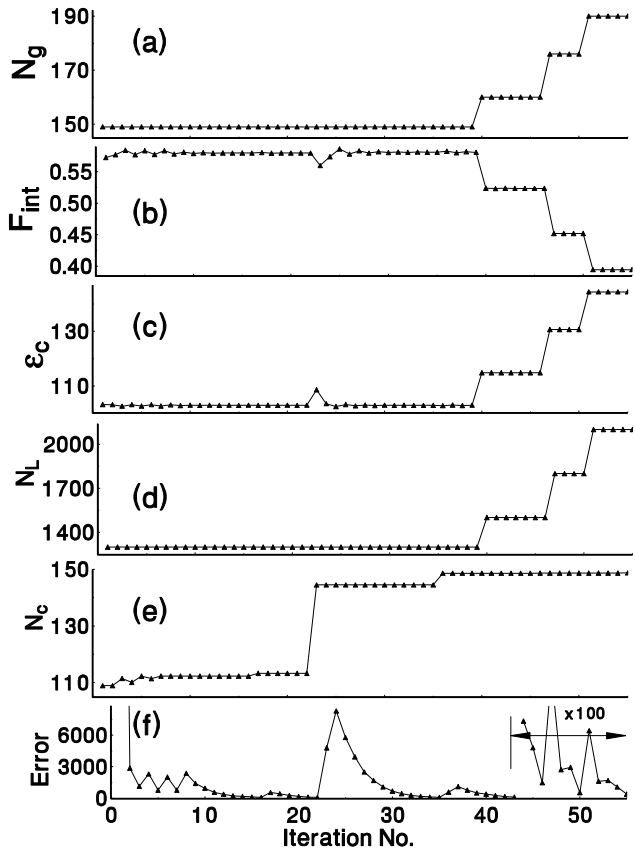,width=\columnwidth,angle=90}
\caption{Convergence of the self-consistency procedure, for $N=2\cdot 10^5$,
$a_{\rm sc}/a_0=0.0072$, $t_{\rm sc}=53$, and a Laguerre DVR mesh. (a) Number
of points in the DVR mesh, $N_g$. (b) Fraction of atoms in the LDA integral,
$F_{\rm int}$. (c) Cut-off energy, $\epsilon_{c}$, specifying the upper limit
of the discrete quasiparticle sum. (d) Order of the Laguerre polynomial,
$N_L$. (e) Condensate number $N_c$. (f) Error, defined by Eq. (\ref{Error}),
showing convergence up to each change of $N_{c}$ or $N_{g}$, and ultimately
convergence to the condition that $N_{c}+N_{T} = N$.}
\label{convt}
\end{figure}

The iteration procedure is illustrated in Fig.~\ref{convt}, which tracks
a calculation for $2\cdot 10^5$ atoms and scaled temperature
$t=k_BT/\hbar\omega=53$ (from Eq.~(\ref{Tc}), $t_c^0\approx 54.3$), using the
Laguerre DVR basis. After more than 50 iterations, $N_{c}$ converged from the
initial estimate of 109 to the final
value of 149 atoms (Fig.~1e). Each adjustment of $N_{c}$ (level 2) or
$N_{L}$ (level 3) results in a jump in the error (Fig.~1f), which then
converges again.  In this
calculation, $N_L$ increased from 1300 to 2100 (Fig.~1d), corresponding to an
increase of mesh points (up to $R_{\rm max}$ = 42) from 149 to 190 (Fig.~1a),
an increase in $\epsilon_{c}$ from $102\hbar\omega_0$ to $144\hbar\omega_0$
(Fig.~1c), and a decrease in the fraction of total number of atoms in the LDA
integral from 57\% to 40\% (Fig.~1b).

The fraction of atoms in the LDA integral is negligible only for calculations
at low temperatures with small $N$. Since $T_{c}$ rises as $\sim 0.94N^{1/3}$,
the required number of thermal states rises with $N$ for calculations near
$T_{c}$, and inevitably the LDA integration must include a larger fraction of
atoms. For $N=2\cdot 10^4$, $2\cdot 10^5$m and $10^{6}$, at most 9\%, 38\%,
and 74\% of the atoms were in the integral at temperatures in the vicinity of
$T_c$. Correspondingly, the mesh size $N_g$ required to ensure convergence
increased from 140 to 210 for $N$ between $10^{3}$ and $10^{6}$. The reason
$N_g$ does not increase more rapidly with $N$ is that the LDA approximation
improves with the total number of atoms.

It should be pointed out that for large values of $N$, the iteration procedure
could exhibit instabilities when the temperature approached $T_{c}$. For
$N>10^5$, we found that there often appeared to be (at least) two semi-stable
regions when $N_{c}\lesssim 5,000$, between which the calculation tended to
fluctuate. In order to ensure the solution remained in the more stable state,
small temperature increments $\Delta t = 0.2$  were used. 

\section{Thermal Averages}
\label{sec:results}

\subsection{Condensate fraction}

In several of the plots to follow, results are presented for a series of
values of $a_{\rm sc}/a_{0}$. For comparison with current experiments, we
note that the scattering lengths $a_{\rm sc}$ for $^{87}$Rb, $^{23}$Na, and
$^7$Li are approximately given by $110a_{B}$, $52a_{B}$, and $-27.3a_{B}$,
respectively, where $a_B\approx 5.292\cdot 10^{-11}$~m is the Bohr radius.
Thus, if one takes $\omega = (\omega_x \omega_y \omega_z)^{1/3}$, then for
the recent MIT experiments~\cite{Ketterle} on $^{23}$Na,
$\nu=\omega/2\pi = 96.4$~Hz, the JILA experiments~\cite{Jin} give
$\nu = 182.5$~Hz, and the Rice experiments~\cite{Bradley} give $\nu =
144.6$~Hz, corresponding to $a_{sc}/a_{0} = 0.00129$, $0.00729$, and
$-0.00046$, respectively. 

\begin{figure}
\psfig{figure=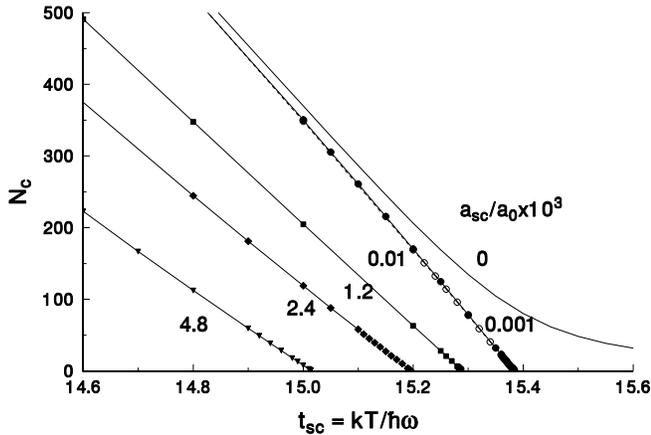,width=\columnwidth,angle=0}
\caption{If $\mu=\tilde{\mu}$, from the HFBP discrete quasiparticle sum (DQS)
near $T_c$ there is a discontinuity in the $N_{c}$ vs.\ $T$ function with
respect to $a_{\rm sc}$. The figures show $N_c$ vs.\ $T$ for $N=5,000$ atoms,
for several values of $a_{\rm sc}/a_{0}$. Even in the limit of small
$a_{\rm sc}/a_{0}$, $N_{c}$ goes to zero abruptly with $T$ for the
self-consistent solution, while for the ideal Bose gas ($a_{sc}=0$),
$N_{c}(T)$ has a smooth tail.}
\label{oldmu}
\end{figure}

Figure \ref{oldmu} illustrates the consequences of setting the eigenvalue of 
the GP equation $\tilde{\mu}$ equal to the chemical potential $\mu$,
as discussed in Section \ref{sec:mft}. With this assumption (here used in
conjunction with the Popov approximation, $m_{T} = 0$), $N_{c}$ goes to zero
abruptly with $T$ when the population in excited states reaches the total
number of atoms $N=5,000$. By contrast, results for $a_{sc}=0$, obtained
as described in Sec. IIE, have a smooth tail at high temperature. Thus,
in the limit $a_{\rm sc} \rightarrow 0$, the results near $T_{c}$ exhibit
a discontinuity with respect to the ideal gas results.

\begin{figure}
\psfig{figure=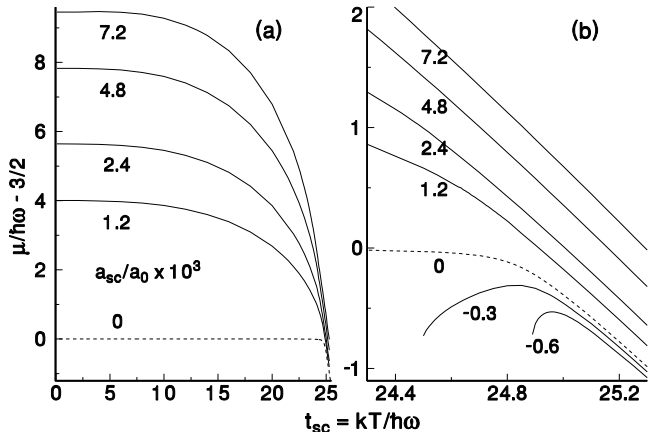,width=\columnwidth,angle=0}
\caption{The chemical potential in units of $\hbar \omega$ relative to the
harmonic oscillator zero-point energy, $\mu/\hbar\omega-3/2$, {\it vs.}
$T$ for various values of $a_{\rm sc}/a_{0}$. (a) Shows the full range
of temperatures up to $T_{c}$, while (b) shows a limited range near $T_{c}$.}
\label{chemt}
\end{figure}

\begin{figure}
\psfig{figure=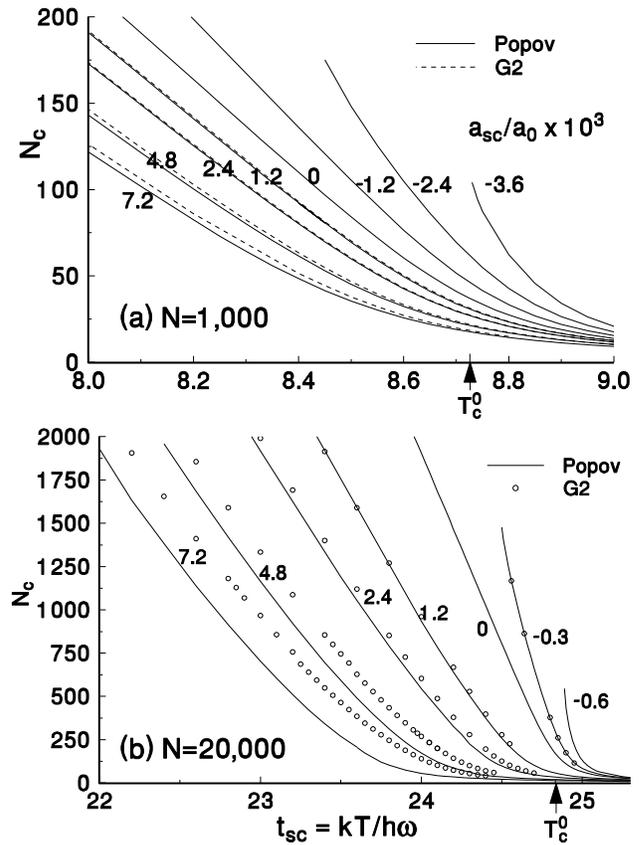,width=\columnwidth,angle=90}
\caption{When $\mu$ differs from $\tilde{\mu}$ according to Eq.~(\ref{muT}),
the $N_{c}$ vs.\ $T$ function from the
discrete quasiparticle sum behaves smoothly with respect to $a_{\rm sc}$.
Shown are the results for (a) $N=1,000$ and (b) $N=20,000$ atoms. The critical
temperature for $a_{\rm sc}=0$, defined as the maximum of $d^2N_c/dT^2$, is
indicated with an arrow. For $a_{\rm sc}<0$ the maximum value of $N_{c}$ is
limited due to the instability of the condensate. In (b), open circles denote
results obtained with the G2 approach.}
\label{newmu}
\end{figure}

Figures~\ref{chemt} and \ref{newmu} show results obtained from calculations
in which the chemical potential is as given in Eq.~(\ref{muT}).
The smooth variation of the chemical potential, Fig.~\ref{chemt},
through $T_{c}$ is reflected in all relevant properties
of the system, including the number of condensate atoms and excitation
frequencies. When $a_{\rm sc}>0$, the chemical potential evolves
continuously from positive to negative values, relative to the harmonic
oscillator zero point energy ${\case 3/2}\hbar\omega$, as the temperature
increases. Since $\mu$ increases with the interaction strength, the value at
which the chemical potential passes through zero increases with $a_{\rm sc}$
even though $T_{c}$ decreases. In addition, Fig.~\ref{chemt} shows that for
$a_{\rm sc}<0$, $\mu<3\hbar\omega/2$ everywhere, with maximum values at
temperatures $T\sim T_c$. 

Figure~\ref{newmu} shows the number of condensate atoms as a function
of temperature for $N=1,000$ and $20,000$ for a range of
interaction strengths $a_{\rm sc}/a_0$, calculated within the DQS formalism.
The condensate population near $T_{c}$ is evidently a continuous function of
both the scattering length and temperature.

The plots shown in Fig.~\ref{newmu}, especially for $20,000$ atoms, show that
the G2 renormalization procedure results in a significantly higher value of
$N_{c}$, relative to that obtained within the Popov approximation, for the
larger values of $a_{\rm sc}$. Furthermore, the difference between the G2 and
Popov results becomes more pronounced as $a_{\rm sc}$ increases. This behavior
is consistent with expectation because G2 produces a weakening of the
atom-atom interaction. The use of the occupation factors (\ref{occ}) rather
than (\ref{Nj}) also increases the value of $N_c$ by a few atoms at high
temperatures, but the effect is much smaller than what results from the
use of G2 theory.

For $a_{\rm sc} < 0$, the $N_{c}$ values reach a maximum when the calculation
becomes numerically unstable \cite{Ruprecht,Houbiers,Hutch,TB}, reflecting the
physical instability of the cloud towards spatial collapse. The maximum
$N_{c}$ values depend on $a_{\rm sc}$, as shown by the termination of the
curves for these cases.  For $T=0$, the maximum value is given by
$N_c^{\rm max}=0.573a_{0}/a_{sc}$~\cite{Ruprecht}. This critical number is
known to decrease when $T>0$ due to the presence of thermal
atoms~\cite{Houbiers,Hutch}. In these plots, the maximum $N_{c}$ is 80\% to
57\% of the value calculated for $T=0$, confirming that the thermal cloud
significantly decreases the stability of the condensate for $a_{\rm sc}<0$.

\subsection{Comparison with LDA and QMC}

It is interesting to explore how our finite temperature results compare with
those obtained by other methods. Local density approximations are much simpler
to implement numerically than the full self-consistent HFB equations and their
variants. The opposite is true of Monte Carlo calculations, but these do not
invoke the mean-field approximation so yield results for equilibrium
configurations that are essentially exact.

\begin{figure}
\psfig{figure=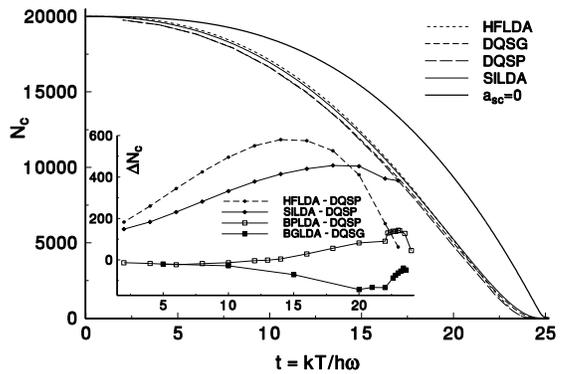,width=\columnwidth,angle=0}
\caption{Comparison of values for $N_{c}/N$ from quasiparticle sums
with the Popov and G2 approximations, as compared with HFLDA and SILDA
for $N=20,000$ atoms and $a_{\rm sc}/a_0=0.0072$. BLDA results are too close
to distinguish on this scale. On an expanded scale, the inset gives
differences between indicated LDA and DQS methods.}
\label{NcLDA}
\end{figure}

Fig.~\ref{NcLDA} for $N=2\cdot 10^4$ compares $N_c$ values from the Popov and
G2 quasiparticle sums (DQSP and DQSG) with severak LDA methods. Our
Hartree-Fock
LDA (HFLDA) solves the GP equation for the condensate $n_{c}(r)$, iterated to
self-consistency using Eq.~(\ref{HFLDA}) for the thermal distribution
$n_{T}(r)$. We found it most efficient to start at low temperature, in order
to obtain good initial estimates of $n_{T}(r)$ at successively higher values
of $T$. No solution could be found for $N_c/N < 0.035$
due to the failure of the HFLDA, as discussed above and in
Ref.~\cite{Holzmann}.

 The `semi-ideal' LDA (SILDA)~\cite{NSK} omits the $n_{T}(r)$ term in the
TF expressions for the condensate~(\ref{TFA}) and for the total density
$n_{T}(r)$ in Eq.~(\ref{HFLDA}). This results in the simple
expressions~(\ref{SILDA}) which are related solely through the chemical
potential. Iterative solution of these equations yields results that are close
to the other functions plotted in Fig.~\ref{NcLDA}. The actual $n_{T}(r)$
distribution calculated with this approach exhibits a sharp peak at edge of
the condensate due to the discontinuity at the Thomas-Fermi condensate radius.

The inset of Fig.~\ref{NcLDA} shows that the Hartree-Fock Bogoliubov LDA
methods, BPLDA and BGLDA, agree most closely with the hybrid method,
DQSP and DQSG, respectively. The two BLDA methods employ a TF condensate,
and thus the $n_{T}({\bf r})$ functions exhibit a small spike at the edge of
the condensate, which has a cusp. As with the HFLDA, the calculations required
iteration to self-consistency, which was facilitated when initial values were
obtained by extrapolation from results from lower temperature values.

It is remarkable that the values for $N_{c}$ from BLDA calculations agreed
with the corresponding DQS results to better than 0.4\% of $N$ in every case
for which results were obtained. Even for HFLDA and SILDA, the differences
with DQS results are less than the fractional error in current experiments.
Thus these comparisons show that relatively simple LDA expressions are useful
for obtaining the condensate fraction as a function of temperature. It is only
in the region near $T_{c}$ and above, where the condensate number becomes
small, that our LDA methods failed.

The Quantum Monte Carlo (QMC) approach uses the exact Hamiltonian with a
hard-sphere atom-atom interaction. Based on extensive numerical experience
with $^{4}$He~\cite{Ceperley}, QMC should be most useful for the calculation of
equilibrium quantities, such as the condensate fraction. Holzmann
{\it et al.}~\cite{Holzmann} have provided benchmark QMC calculations for the
case of $10^4$ Bose atoms confined in a spherical trap, with
$a_{\rm sc}/a_{0} = 0.0043$.  Table~\ref{table} shows comparisons between our
results and those of QMC~\cite{Holzmann,H2} for the condensate number as a
function of temperature. The DQSP, DQSG, and QMC values differ by up to 1.2\%
of the total atom number $N$. It is notable that at higher temperatures,
$N_{c}$ falls off less quickly using HFBP and G2 than QMC. This may be due in
part to the fact that the relationship between $N_{c}$ and $\mu$ in
Eq.~(\ref{muT}) is not entirely correct at higher temperatures, as discussed
in Section~\ref{sec:mu}, and may resemble ideal gas statistics too closely.
Presumably the many-body effects that necessitate the renormalization of the
atom-atom interaction are already included in the QMC procedure, in which case
results with G2 should be closer than Popov to the QMC. Indeed, for
$t = k_{B}T/\hbar \omega < 17$, the Popov results lie below QMC, while the G2
numbers are higher and closer to QMC. Above a scaled temperature $t = 18$,
however, the G2 results rise above QMC values.

\subsection{Critical Temperature vs. $a_{\rm sc}$}

Figures~\ref{newmu} and \ref{NcLDA} show that large values of $a_{\rm sc}$
have the effect of flattening the curve of condensate number as a function of
temperature, as is already apparent in the plots of Giorgini
{\it et al.}~\cite{GiorginiLTP}. If these curves are fit to a function
$N_{c}/N = 1 - (T/T_{c})^{\alpha}$, one obtains values for $\alpha$ as low as
$1.4$, compared with the ideal gas value of $3$. Another parameter to
characterize the effect of atom-atom interactions is the shift of the critical
temperature from the ideal Bose gas case. For the homogeneous Bose gas, where
it is uniquely defined as the point at which $N_{c}$ goes to zero, this shift
has been the subject of intense discussion recently~\cite{HUH}. For atoms in a
harmonic potential, as is especially clear in Fig.~\ref{newmu}, this point is
not sharp (indeed, the number of condensate atoms is finite at all temperatures
in a mesoscopic system). Definitions of $T_{c}$ that have been proposed include
the point at which the density at the origin reaches the critical density
for a homogeneous gas~\cite{StoofTc}, the maximum of the specific heat, and
the maximum of the temperature derivative of the specific heat~\cite{Balazs}.
Since such energy-weighted properties pose additional problems for numerical
calculations of thermal averages, $T_{c}$ is determined here as the maximum of
the function $d^2N_c/dT^2$. The inflection point of the $N_{c}$ {\it vs}.\ $T$
function, or zero of $d^{2}N_{c}/dT^{2}$ deviated from Eq. (\ref{Tc}) by a
significantly larger amount.

\begin{figure}
\psfig{figure=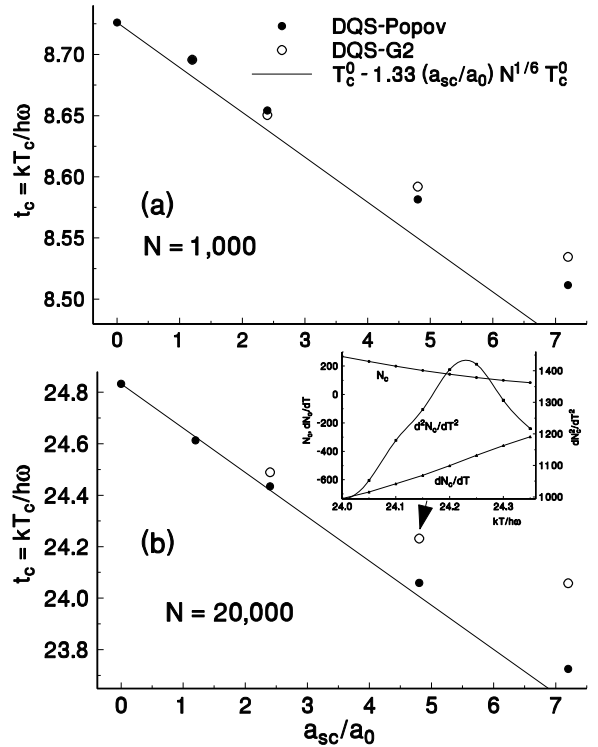,width=\columnwidth,angle=90}
\caption{Values for the critical temperature, $T_{c}$, defined as the
maximum of $d^{2}N_{c}/dT^{2}$. Results are shown for (a) $N=1,000$ and
(b) $N=20,000$ atoms, for the DQSP and DQSG approaches. The solid line
represents the semiclassical prediction $T_c=T_{c}^{0} - \Delta T_{c}$, where
$T_{c}^{0}$ is the transition temperature in the non-interacting limit. The
inset in (b) shows the $N_{c}(T)$, $N_{c}(T)/dT$, and $d^{2}N_{c}/dT^{2}$
functions from which $T_{c}$ is determined for the cases
$a_{\rm sc}/a_0=0.0048$, the last with a spline fit.}
\label{tctn}
\end{figure}

Figure~\ref{tctn} shows $T_{c}$ values extracted from the data used in
Fig.~\ref{newmu}. For comparison, the ideal gas data are analyzed in a similar
manner, yielding values of $T_{c}$ that are close to, but not identical with,
those obtained using Eq.~(\ref{Tc}). Figures \ref{tctn}a and \ref{tctn}b
correspond to $1,000$ and $20,000$ atoms, respectively. The inset in
Fig.~\ref{tctn}b shows how the transition temperature is determined from the
data in a typical case, by making use of the three functions $N_{c}(T)$,
$dN_{c}(T)/dT$, and $d^{2}N_{c}/dT^{2}$. Since both the condensate number and
its temperature derivative are nearly straight lines, accurate calculation of
the second derivative requires accurate numeric values of these functions.

A semiclassical analysis by Giorgini et al.~\cite{GiorginiTc} indicates that
the transition temperature should decrease linearly from the ideal gas value
with increasing particle interactions. The results of the DQS-Popov
calculations confirm this general scaling; furthermore, as the number of atoms
increases, the observed shift in the critical temperature $\delta T_c$ matches
the semiclassical expression more closely at larger $a_{\rm sc}/a_0$. In
contrast, with the DQS-G2 approach $\delta T_{c}$ shows significant deviations
from linear scaling for small $N$, and these become more pronounced as the
number of atoms increases. For $N=2\cdot 10^4$, the shifts are significantly
less than the semiclassical values for the larger values of $a_{\rm sc}/a_{0}$
considered.

\subsection{Renormalization of the atom-atom interaction}

As indicated in Fig.~\ref{newmu} above, the G2 renormalization yields values
for $N_{c}/N$ that reflect the weakening of the atom-atom repulsion; at any
given temperature, the number of atoms in the condensate increases relative to
the value obtained using the Popov approximation. Perhaps more interesting is
the spatial variation of the effective interaction in the harmonic trap. The
renormalization is governed by the local value of $m_T$ relative to $n_c$. In
general, $|m_T|$ increases with the number of noncondensed atoms $n_T$ since
more terms enter the sum~(\ref{anomaly}); however, $m_T$ vanishes when
$n_c=0$, since the `quasihole' amplitude $v_i=0$. In general, therefore, one
might expect the local renormalized interaction to reach a minimum at some
temperature. For a uniform Bose gas, this minimum occurs at exactly the
transition temperature, and corresponds to a vanishing of the effective
scattering length~\cite{BStoofRG,ShiGriffin}.

\begin{figure}
\psfig{figure=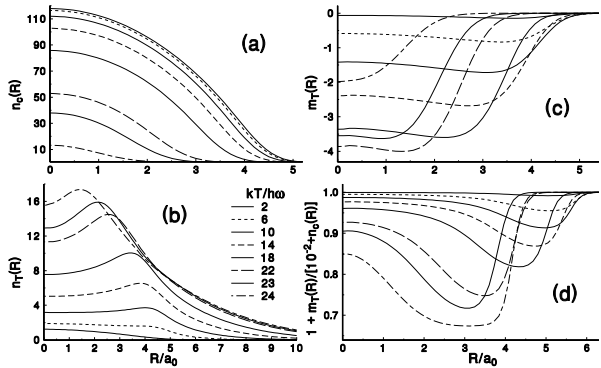,width=\columnwidth,angle=0}
\caption{The functions $n_{c}(r)$, $n_{T}(r)$, $m_{T}(r)$, and $g(r)/g$ is 
shown for $N=20,000$ atoms, $a_{\rm sc}/a_{0}=0.0072$ over a wide range of
temperatures. $|m_{T}(r)|$ is largest at the edge of the condensate and
increases with $T$ up to $T_{c}$.}
\label{mT20k}
\end{figure}

In Fig.~\ref{mT20k} we compare the condensate and thermal densities with the
spatial variations of the anomalous average and the effective particle
interactions for the case of $N = 20,000$ and $a_{\rm sc}/a_{0} = 0.0072$ for
various temperatures. (As noted above, this would correspond to a relatively
tight trap for $^{23}$Na.) For these plots, $\delta=0.01$ in Eq.~(\ref{gr}).
There
is a slight dependence of the results on $\delta$, since much smaller values
$\delta\sim 10^{-4}$ lead to a small bump in the $m_{T}(r)$ function at the
very edge of the condensate. The dependence on this arbitrary parameter
indicates an ambiguity in the theory; however, the integrated numbers are not
significantly altered by the choice of $\delta$, since the errors are incurred
in regions of very small condensate density.

\begin{figure}
\psfig{figure=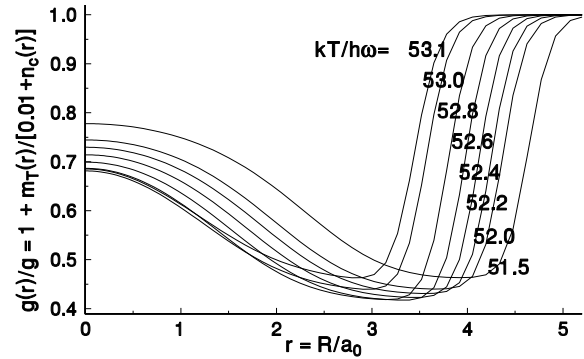,width=\columnwidth,angle=0}
\caption{Variation of the renormalization factor, $g(r)/g$ with temperature
near $T_{c}$ for $N$ = 200,000, and $a_{\rm sc}/a_{0} = 0.0072$. The range of
the minimum decreases as the condensate shrinks with $T$, while the minimum
value continues to decrease up to a point, and then increases.}
\label{mT200k}
\end{figure}

\begin{figure}
\psfig{figure=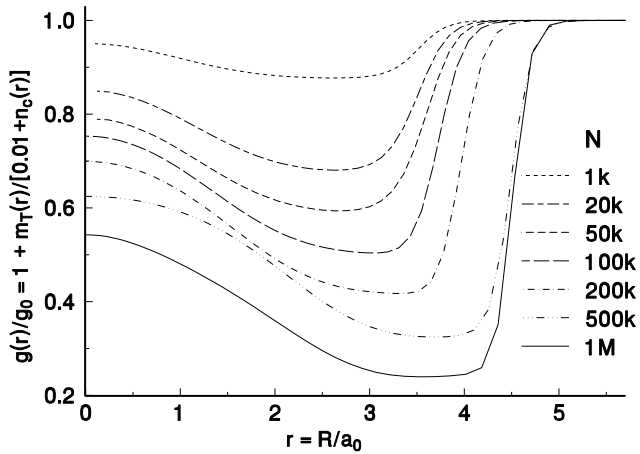,width=\columnwidth,angle=0}
\caption{The curves shown are the `minimum' functions, $g_{\rm min}(r)/g$,
as a function of temperature (such as shown in the previous figure) for each
$N$ value given. These curves are for $a_{\rm sc}/a_{0} = 0.0072$.}
\label{grmin}
\end{figure}

The manner in which $g(r)/g$ attains a minimum in $r$ is shown in
Fig.~\ref{mT200k} for the particular case of $N = 2\cdot 10^5$. The global
minimum occurs at a temperature close to $T_c$, defined above. Following
this procedure, we consider the $g_{\rm min}(r)/g$ functions for various
values of $N$ for $a_{\rm sc}/a_0=0.0072$, which are displayed in
Fig.~\ref{grmin}. Though we have increased $N$ without changing the trap
frequency, the approach to the thermodynamic limit is beginning to emerge. The
minimum for each $N$ is found to always occur very close to the calculated
transition temperature, and its value decreases approximately with $\log(N)$
over the range of $N$ considered. For $N=10^6$, we obtain
$g_{\rm min}(r)/g\approx 0.2$. It should be noted that although the fraction
of total atoms in the LDA integral increased to approximately ${\case3/4}$ for
$N = 10^{6}$ near $T_c$, the high-energy LDA contribution to $m_{T}$ was in
every case less than 2\%, and typically an order of magnitude less than this
value. 

It should be emphasized that the G2 renormalization employed in the present
calculations is derived for a uniform Bose gas, and should best represent
large condensate densities or low temperatures where the LDA is most
applicable. While the LDA is bound to fail for $T\to T_c$, the regime
where it loses validity will become smaller with increasing $N$, and should
approach the critical region where perturbation theory itself breaks
down. It would be preferable to define the renormalization of the particle
interactions in terms of the full many-body T-matrix in a trap, and we hope to
pursue this issue in future work. The G2 approach as formulated above, however,
should properly describe the effects of two-body correlations for large
trapped condensates at low to intermediate temperatures. Thus, the strong
reduction in the effective interaction strength over much of the condensate,
indicated by the G2 theory, could have significant experimental consequences.
The predictions for the excitation frequencies are discussed further below.

\subsection{Excitation frequencies}

The quasiparticle eigenvalues correspond to excitation frequencies, but it
remains unclear what relationship exists between these values and
experimentally observed resonances of the trapped gas at finite temperatures
when the potential of a harmonic trap is perturbed periodically. In all
mean-field calculations such as those presented here, the linear response
equations assume that the thermal density is fixed, while in experiments it
would also be perturbed. For this reason, the dipole excitation frequency
obtained within mean-field theories will generally not satisfy the generalized
Kohn theorem~\cite{Dobson}, which states that there is a mode in which the
entire ensemble oscillates at the bare trap frequency. Calculations explicitly
including the dynamics of both $n_c$ and $n_T$ \cite{ZGN,BijlsmaRes} are found
to be consistent with the Kohn theorem.

\begin{figure}
\psfig{figure=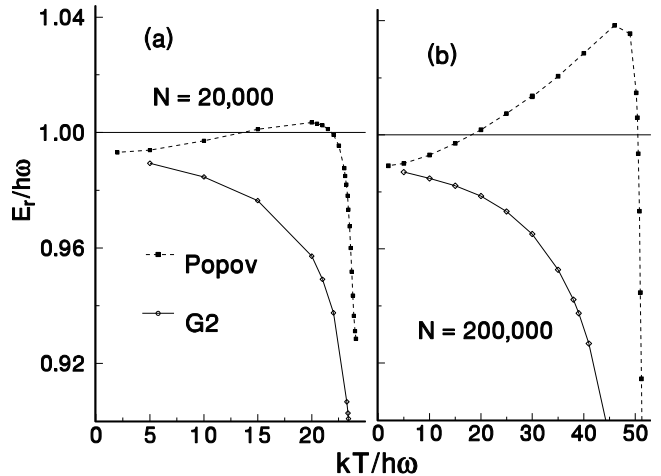,width=\columnwidth,angle=0}
\caption{Excitation frequencies of the lowest $\ell=1$ mode in comparison
with the Kohn theorem value of unity. Results from the Popov (dashed lines)
and G2 (solid lines) approximations are shown for (a) $N=2\cdot 10^4$ and (b)
$N=2\cdot 10^5$. All results are for $a_{\rm sc}/a_{0}=0.0072$.}
\label{Kohn}
\end{figure}

Figure~\ref{Kohn} shows small but significant deviations in the Kohn mode
from unity for $N=2\cdot 10^4$ and $2\cdot 10^5$, both within DQS-Popov and
DQS-G2. That the G2 frequency should be lower than the Popov value cannot be
simply understood in terms of an overall decrease in the interatomic repulsion,
since this would predict a mode closer to unity. Rather, the spatial variation
of the effective interaction leads to a flattening of the effective potential,
comprised of the trap plus the Hartree potential; the looser effective
confinement softens all the modes. We are not aware of other computational
results in which the Popov value starts from below unity and rises above,
before falling near $T_{c}$. This behavior may be a consequence of a more
rigorous treatment of the chemical potential, Eq.~(\ref{muT}). Alternatively,
since the differences increase with $N$ (specifically, the non-condensate
density), they may not have been observable with the smaller $N$ values
studied previously.

\begin{figure}
\psfig{figure=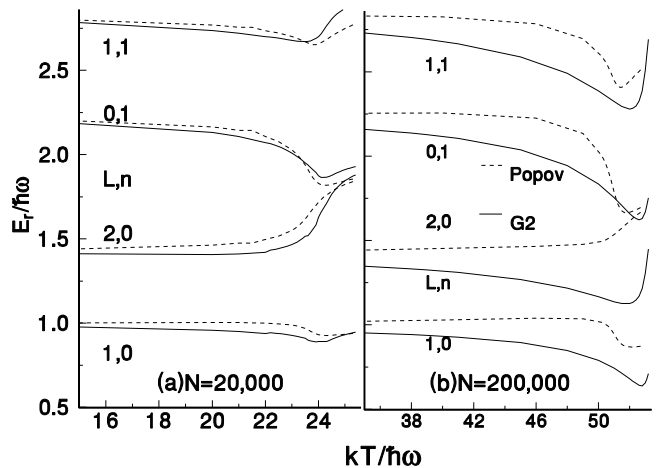,width=\columnwidth,angle=0}
\caption{Excitation frequencies for the lowest $\ell=0$, 1, and 2 modes
for (a) $N=2\cdot 10^4$ and (b) $N=2\cdot 10^5$ within the Popov (dashed
lines) and G2 (solid lines) models, where $a_{\rm sc}/a_{0}=0.0072$.}
\label{frq}
\end{figure}

The temperature-dependence of the low-lying excitation frequencies obtained
with the DQSP and DQSG approaches is shown in Fig.~\ref{frq} for
$N=2\cdot 10^4$ and $2\cdot 10^5$. The softening of all the excitation
frequencies in the G2 approximation was found previously by the proponents of 
this theory~\cite{HDB} (for a `pancake' geometry) as well as by others using
a similar perturbative approach to the interacting Bose gas~\cite{Reidl2}.
However, for a spherically symmetric trap, the results of Ref.~\cite{HDB} for
2,000 Rb atoms showed only a negligible difference between Popov and G2
excitation frequencies. The present results show that for a spherically
symmetric trap and larger atom numbers, there can be differences between the
Popov and G2 values that would be experimentally detectable. These results
also lead to the question whether for larger atom numbers, a renormalized
atom-atom interaction would effect frequencies calculated by the methods of
Refs.~\onlinecite{ZGN,BijlsmaRes}, which did {\it not} assume a static
condensate. It should also be mentioned that experimentally observed
excitation frequencies with larger numbers of sodium atoms in a `cigar'
geometry~\cite{Ketterle} also exhibited a softening of both the quadrupole and
{\it dipole} excitation frequencies as the temperature approaches $T_{c}$. 

\section{discussion and conclusions}
\label{sec:conclusions}

In this work, we have extended finite-temperature mean-field calculations for
Bose-Einstein condensates confined in harmonic traps \cite{HZG,HDB}. A
careful derivation of
the mean-field equations provides improved definitions of the thermodynamic
chemical potential and quasiparticle occupation factors, yielding observables
that are continuous functions of the particle interactions. The numerical
techniques employed in the calculations have allowed for the investigation of
systems with the large numbers of atoms relevant to on-going experiments. In
the process, we have been able to make several crucial comparisons between the
results of evaluating discrete summations over quasiparticle states (which are
numerically time-consuming) and various local density approximations.
Furthermore, we have explored the implications of a recently proposed gapless
theory which takes into account pairing correlations.

The results presented above indicate a significant inadequacy of
conventional static mean-field theory for computations of excitation
frequencies of trapped Bose condensates at finite temperatures.
For large number of atoms and interaction strength, we find appreciable
deviations of the dipole frequency obtained with either the Popov or G2
approximations from expectations of the generalized Kohn's theorem.
In our computations, the condensate is static in the presence of
thermal excitations. The excited dipole mode corresponds approximately to
out-of-phase motion of the thermal cloud relative to the condensate,
as observed experimentally \cite{Ketterle} when the dipole
mode of the thermal cloud is excited separately. Detailed modelling of
such excitation modes has been performed only by restrictive
parametrization of the condensate and thermal cloud in the
collisionless~\cite{BijlsmaRes} or hydrodynamic~\cite{ZGN} regimes. Both of
these approaches address the two-fluid nature of these systems, and 
produce dipole modes that satisfy the Kohn theorem exactly. We will argue
that equilibrium thermal excitations are computed accurately by the
mean-field DQS methods presented here.  However, any experimental probe
of these excitations involves perturbative processes that require other
theoretical methods.

In principle, mean-field theories that include fluctuations in the population
of excited states~\cite{Proukakis,Giorgini98} ought to be equivalent to the
two-fluid dynamics in the collisionless regime. A full second-order
perturbation theory of the interacting Bose gas should yield the coupled
modes of the condensate and thermal clouds as well as damping rates. Indeed,
employing the approximate many-body T-matrix in the calculations (the G2 
approximation described above) yields excitations that have a
temperature-dependence qualitatively similar to that of out-of-phase modes. We
hope to explore these issues in future work.

\begin{acknowledgments}

This work was supported by the National Science Foundation (TB and BIS), the
Office of Naval Research (TB and DLF), and by a grant from the NCF-Cray
Foundation of the Netherlands (TB). The authors are grateful for valuable
conversations with H.~Beijerinck, K.~Burnett, C.~W.~Clark, K.~K.~Das, M.~Doery,
M.~Edwards, M.~Gajda, A.~Griffin, D.~A.~W.~Hutchinson, H.~Metcalf, H.~Stoof,
E.~Vredenbregt, and E.~Zaremba. The authors particularly appreciate
S.~A.~Morgan for his valuable comments and for sending us a preliminary draft
of his D.~Phil.\ thesis.
\end{acknowledgments}

\begin{table}
\caption{Comparison of condensate numbers, $n_{0}$, obtained from
Quantum Monte Carlo calculations
and from this work, with and without atom-atom
interactions, and results obtained here from discrete Bogoliubov
quasiparticle sums and discrete Hartree-Fock sums. The error limits for QMC
are of course purely positive for $n_{0}=0$.}
\begin{tabular}{lrrrrr}
 & QMC & DQSG & DQSP & HFLDA\tablenote{Holzmann et al.~\cite{Holzmann,H2}} &
HFLDA\tablenote{This work, using Eqs.~(\ref{SILDA}).} \\
$T/\hbar\omega$ & $N_{c}$ &$N_{c}$ & $N_c$ &$N_c$ &$N_c$  \\
16.667  &  2265(10) & 2213 & 2159 &  2216  &  2222  \\
16.949  &  1971(10) & 1936 & 1883 & 1945  &  1935  \\
17.242  &  1656(15) & 1654 & 1599 & 1630  &  1638   \\
17.544  &  1374(10) & 1367 & 1309 & 1323  &  1333   \\
17.857  &  1057(10) & 1072 & 1016 & 1008  &  1022  \\
18.182  &   741(10) &  782 &  726 & 686  &      \\
18.519  &   440(10) &  501 &  448 &     &   \\
18.868  &   180(10) &  247 &  205 &    &     \\
19.231  &    21(11) &  140 &   57 &    &       \\
19.608  &     0(20) &   71 &   21    &  &     \\
19.802  &     0(20) &       &  15 &    &      \\
20.0    &     0(10) &       & 12 &    &       \\
20.202  &     0(14) &       &   9 &    &     \\
\end{tabular}
\label{table}
\end{table}

\end{document}